\documentclass[aps, pre, twocolumn, amsmath, amssymb, superscriptaddress, longbibliography]{revtex4-1}
\usepackage{graphicx}
\usepackage{dcolumn}
\usepackage{bm}
\usepackage[dvipsnames]{xcolor}
\usepackage{siunitx}

\definecolor{darkblue}{rgb}{0,0,0.6}
\definecolor{darkred}{rgb}{0.6,0,0}
\usepackage[colorlinks=true,urlcolor=darkblue, citecolor=darkblue, linkcolor=darkred, hyperfootnotes=false]{hyperref}

\DeclareSIQualifier{\pp}{pp}

\usepackage[textsize=tiny, color=green!40]{todonotes}

\newcommand{\mcC}{\mathcal{C}}

\newcommand{\ee}{\boldsymbol{e}}

\newcommand{\kk}{\boldsymbol{k}}
\newcommand{\Pe}{\mathrm{Pe}}
\newcommand{\rr}{\boldsymbol{r}}
\newcommand{\mcR}{\mathcal{R}}

\newcommand{\xx}{\boldsymbol{x}}

\newcommand{\eeta}{\boldsymbol{\eta}}

\def \equi#1{\mathrel{\mathop{\kern 0pt\sim}\limits_{#1}}}

\newcommand{\ratio}{:}

\renewcommand\vec{\mathbf}

\graphicspath{{../Figures/}}

\begin{document}

\title{Pair correlation of dilute Active Brownian Particles: from low activity dipolar correction to high activity algebraic depletion wings}

\date{\today}
\author{Alexis Poncet}
\affiliation{LPTMC, CNRS/Sorbonne Universit\'e, 4 Place Jussieu, F-75005 Paris, France}
\affiliation{Department of Physics, The University of Tokyo, Hongo 7-3-1, Tokyo, 113-0033, Japan}

\author{Olivier B\'enichou}
\affiliation{LPTMC, CNRS/Sorbonne Universit\'e, 4 Place Jussieu, F-75005 Paris, France}

\author{Vincent D\'emery}
\email{vincent.demery@espci.psl.eu}
\affiliation{Gulliver, CNRS, ESPCI Paris, PSL Research University, 10 rue Vauquelin, Paris, France}
\affiliation{Univ Lyon, ENS de Lyon, Univ Claude Bernard Lyon 1,
CNRS, Laboratoire de Physique, F-69342 Lyon, France}

\author{Daiki Nishiguchi}
\affiliation{Department of Physics, The University of Tokyo, Hongo 7-3-1, Tokyo, 113-0033, Japan}

\begin{abstract}
We study the pair correlation of Active Brownian Particles at low density using numerical simulations and analytical calculations.
We observe a winged pair correlation: while particles accumulate in front of an active particle as expected, the depletion wake consists of two depletion wings.
In the limit of soft particles, we obtain a closed equation for the pair correlation, allowing us to characterize the depletion wings.
In particular, we unveil two regimes at high activity where the wings adopt a self-similar profile and decay algebraically.
We also perform experiments of self-propelled Janus particles and indeed observe the depletion wings.
\end{abstract}

\maketitle

\section{Introduction}\label{}

The pair correlation function has played a pivotal role in our understanding of the structure of equilibrium liquids~\cite{Hansen2006}; the same is to be expected for active liquids, such as bacterial colonies~\cite{Zhang2010,Nishiguchi2017}, flocks of starlings~\cite{Cavagna2010, Cavagna2018} or colloidal rollers~\cite{Bricard2013,Geyer2018,Geyer2019}, or assemblies of Janus particles~\cite{Palacci2010, Ginot2015, Zhang2020, Iwasawa2020}.
It has indeed been used to quantify the order in bacterial colonies~\cite{Zhang2010} and to infer the interactions in bird flocks~\cite{Cavagna2010, Cavagna2018}.
However, its broad utilization is precluded by the lack of analytical results on its general form, notably little is known about its regions of positive and negative sign, even in the homogeneous phase and in the absence of alignment interactions~\cite{Bialke2013, Solon2015, MariniBettoloMarconi2015}.

The theoretical prediction of the pair correlation is indeed an important challenge, which has been undertaken for minimal models without alignment, such as Active Browian or Ornstein-Uhlenbeck particles (ABP or AOUP). 
So far, even for these minimal models, the study of the pair correlation focused mostly on the explanation of the phase separation that occurs at large activity~\cite{Cates2015, Digregorio2018, Klamser2018}.
Two main approaches have been followed.
The first consists in computing the angular average of the pair correlation due to an activity represented by a persistent translational noise. 
An attractive term has been found, and interpreted as a higher tendency to phase separate~\cite{Farage2015, MariniBettoloMarconi2015, Fodor2016b}.
However, this approach does not, by nature, retain the angular dependence of the correlations.
The second approach consists of a quantitative prediction of the polar pair correlation using a closure of the many body Smoluchowski equation leading to the effective velocity, whose decrease with density may explain the phase separation~\cite{Bialke2013, Takatori2014, Solon2015, Hartel2018, Bickmann2020}.
This method requires a numerical solution of nonlinear equations, 
and does not provide explicit predictions.
Finally, up to now there is no analytical characterization of the pair correlation.

In this article, we address the global shape of the pair correlation of ABP in the dilute and homogeneous regime~\cite{Digregorio2018}.
We first observe a winged pair correlation in numerical simulations of ABP: while particles accumulate in front of an active particle as expected, the depletion wake consists of two depletion wings (Fig.~\ref{fig:scheme_examples}).
We then resort to a linearized Dean equation to obtain a closed equation for the pair correlation.
Solving this equation under different limits, we unveil three different regimes for the correlations, which we organize on a phase diagram.
In two regimes, the wings adopt a self-similar profile and decay algebraically.
Last, we measure the pair correlation in experiments of self-propelled Janus particles, and qualitatively observe the depletion wings.

\section{Model and numerical simulations}\label{}

\begin{figure}
	\centering
	\includegraphics[scale=1]{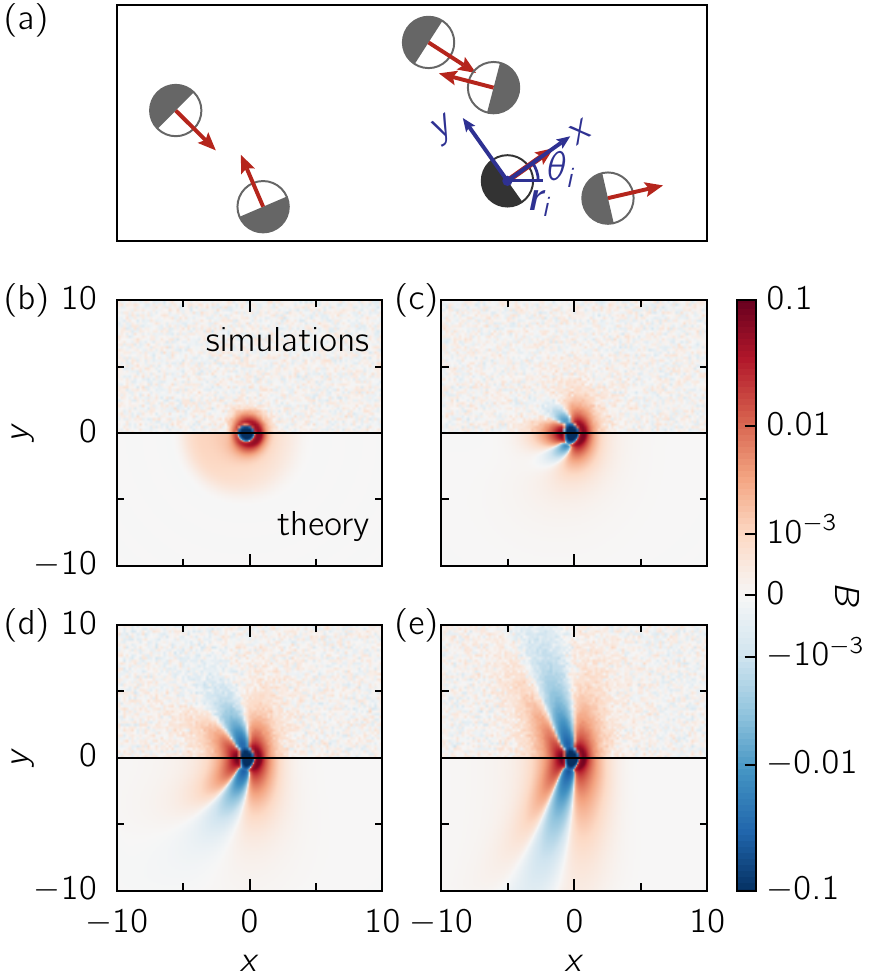}
	\caption{
		(a) Snapshot of numerical simulations of ABP. The red arrows indicate the orientations of the particles. The blue axes represent the local frame used to compute the pair correlations.
		(b)-(e) Evolution of the correlations with decreasing $D_r$. $\phi = 0.04$ and
		$\epsilon\ratio D\ratio U = 1\ratio 0.1\ratio 10$, and $D_r = 10, 1, 0.1, 0.01$ ($\Pe=10, 31.6, 100, 316$).
		Top: simulations. Bottom: numerical integration of Eq.~\eqref{eq:correl_dir}.
	}
	\label{fig:scheme_examples} 
\end{figure}

First, we perform numerical simulations of ABP in dimension two interacting via a pairwise harmonic potential $V(r)=\frac{\epsilon}{2}(1-r/a)^2$ for $r<a$, where $\epsilon$ sets the interaction strength and $a$ is the particle diameter 
(Fig.~\ref{fig:scheme_examples}(a), see App.~\ref{app:sim_desc} for details).
The position $\rr_i(t)$ and orientation $\theta_i(t)$ of the particle $i$ follow
\begin{align}
\dot \rr_i &= -\gamma^{-1}\nabla_i \sum_{j\neq i} V(\rr_i-\rr_j)
 + U \hat \ee_{\theta_i} + \sqrt{2D} \eeta_i, \label{eq:X} \\
\dot\theta_i &= \sqrt{2D_r} \nu_i, \label{eq:theta}
\end{align}
where $U$ is the propulsion velocity, $D$ is the translational diffusion coefficient, $D_r$ is the rotational diffusion coefficient,
$\hat \ee_{\theta}$ is the unit vector with polar angle $\theta$
and $\eeta_i$ and $\nu_i$ are normalized Gaussian white noises.
The diameter $a$ of the particles and the friction coefficient $\gamma$ can be set to one through a rescaling; the parameters $\epsilon$, $U$, $D$ and $D_r$ are given in arbitrary units and only their relative values are important.
We use a low area fraction $\phi\simeq 0.04$ to focus on two-body effects and remain in the homogeneous phase~\cite{Cates2015,Digregorio2018, Klamser2018} (see discussion in App.~\ref{app:sim_extension}).

The polar pair correlation $C(\rr,\theta,\theta')$ is defined from the density field for the positions $\rr_i(t)$ and orientations $\theta_i(t)$,
$f(\rr, \theta, t) = \sum_i \delta(\rr_i(t) - \rr)\delta(\theta_i(t) - \theta)$:
\begin{equation}\label{eq:def_correl}
 C(\rr, \theta, \theta') =
 \frac{\langle f(0, \theta)f(\rr, \theta')\rangle}{[\rho/(2\pi)]^2} - \frac{\delta(\rr)\delta(\theta-\theta')}{\rho/(2\pi)}
 -1,
\end{equation}
where $\rho=4\phi/(\pi a^2)$ is the density;
the correlation of a particle with itself is removed in the second term, and the $r\to\infty$ limit is removed in the third.
From rotational invariance, $C(\rr,0,\theta')$ contains all the information in $C(\rr,\theta,\theta')$.
We focus on the density of particles in the reference frame of a given particle (blue axes in Fig.~\ref{fig:scheme_examples}(a)~\cite{Zhang2010}), which retains the polar character of the correlations:
\begin{equation}\label{eq:def_B}
B(\rr) = \frac{1}{2\pi}\int_0^{2\pi} C(\rr, 0, \theta') d\theta'.
\end{equation}
Note that if translational and rotational degrees of freedom decouple, for instance in absence of propulsion, the isotropic pair correlation is recovered: $C(\rr,\theta,\theta')=B(\rr)=h(r)$~\cite{Hansen2006}. 

The polar pair correlation $B(\rr)$ in the simulations for $\epsilon\ratio D\ratio U = 1\ratio 0.1\ratio 10$, and $D_r = 10, 1, 0.1, 0.01$ is shown in Fig.~\ref{fig:scheme_examples}(b)-(e).
At large rotational diffusion, $D_r=10$, the correlation is nearly axisymmetric and decays quickly (Fig.~\ref{fig:scheme_examples}(b)).
At smaller rotational diffusion, $D_r\leq 1$, as expected, the correlation is positive in front of the particle, indicating an accumulation of other particles.
However, while a depletion wake is expected behind the active particle, as in active microrheology~\cite{Meyer2006, Demery2014c} or in driven binary mixtures~\cite{Poncet2017}, the depletion concentrates in two ``wings'' on the sides of the particle.
As $D_r$ decreases, the characteristic depletion wings appear, their length increases and their curvature decreases.

\begin{figure}
\centering
\includegraphics[scale=.9]{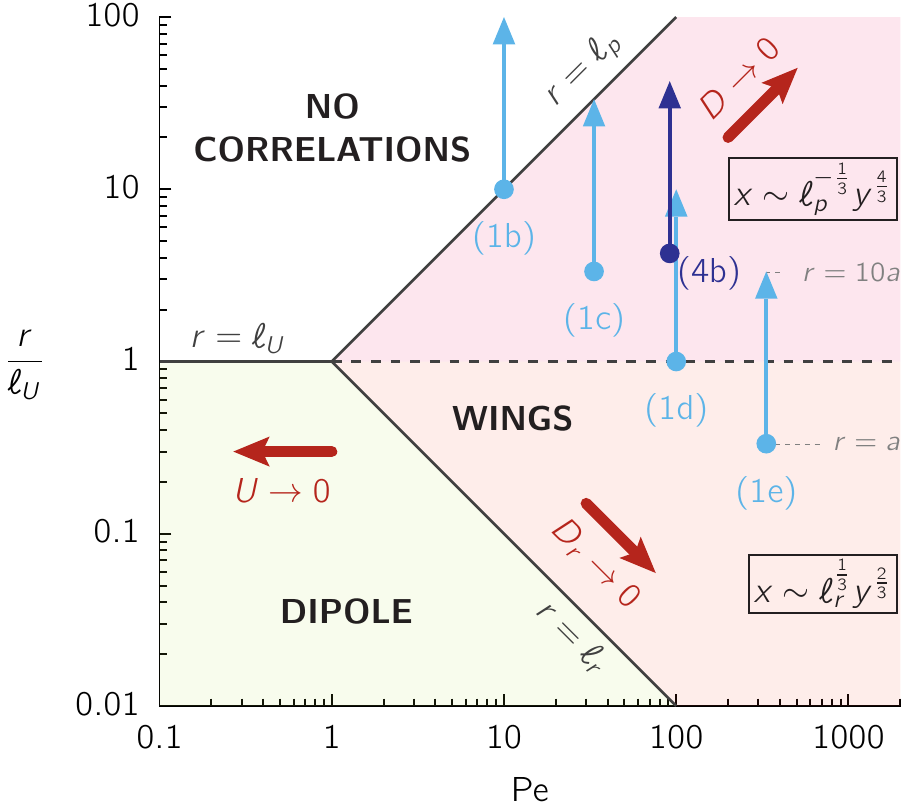}
\caption{
Phase diagram showing the different regimes as a function of P\'eclet number and observation length. 
Blue arrows indicate the regimes explored in the simulations (Fig.~\ref{fig:scheme_examples}(b)-(e)) and in the experiments (Fig.~\ref{fig:experiments}) and , the observation length ranges from the diameter $a$ of the particles to $10a$.
Red arrows point to the limiting regimes explored in Fig.~\ref{fig:small_U}.
}
\label{fig:phase_diag}
\end{figure}

\section{Theory and limiting regimes}\label{}

To rationalize these different regimes, we consider the lengthscales and dimensionless parameters of the problem.
In the dilute limit considered here, the structure of the pair correlation beyond the size of a particle is controlled by the parameters $U$, $D$ and $D_r$, which combine into three lengthscales
\begin{equation}
\ell_r=\frac{D}{U};\quad\ell_U=\sqrt{\frac{D}{D_r}};\quad\ell_p=\frac{U}{D_r},
\end{equation}
whose relative values are set by the P\'eclet number
\begin{equation}
\Pe=\frac{U}{\sqrt{DD_r}}=\frac{\ell_U}{\ell_r}=\frac{\ell_p}{\ell_U}.
\end{equation}
The P\'eclet number varies between $\Pe=10$ and $316$ in 
Fig.~\ref{fig:scheme_examples}(b)-(e).
For spherical particles the coefficients $D$ and $D_r$ are related through $D_r\sim D/a^2$; here we regard them as independent parameters to disentangle the effects of translational and rotational diffusion.
The form of the pair correlation thus depends on the P\'eclet number and the relative value of the observation length $r$ with the three lengthscales, so that the numerical simulations can be placed on a parameter plane (blue arrows in Fig.~\ref{fig:phase_diag}).

To account for the observed correlations and characterize the shape of the depletion wings, we resort to a linearized Dean equation for the density field $f(\rr,\theta,t)$~\cite{Dean1996, Farrell2012, Demery2014c, Dean2014, Demery2016b, Poncet2017}, which is valid for weak interactions.
At low density, we show in App.~\ref{app:th_eq} that the pair correlation satisfies:
\begin{equation} \label{eq:correl_dir}
 \left[2D\nabla^2 + D_r(\partial_\theta^2 + \partial_{\theta'}^2)+U (\hat\ee_{\theta}-\hat\ee_{\theta'})\cdot\nabla\right]
 C
=- \frac{2}{\gamma}\nabla^2 V.
\end{equation}
Solving this equation numerically (App.~\ref{app:numint}), we obtain an excellent agreement with the numerical simulations
(Fig.~\ref{fig:scheme_examples}(b)-(e)).
We now show that this equation captures the structure of the pair correlation by examining its limiting regimes $U\to 0$, $D_r\to 0$, and $D\to 0$.

\begin{figure*}
\centering
\includegraphics[scale=1]{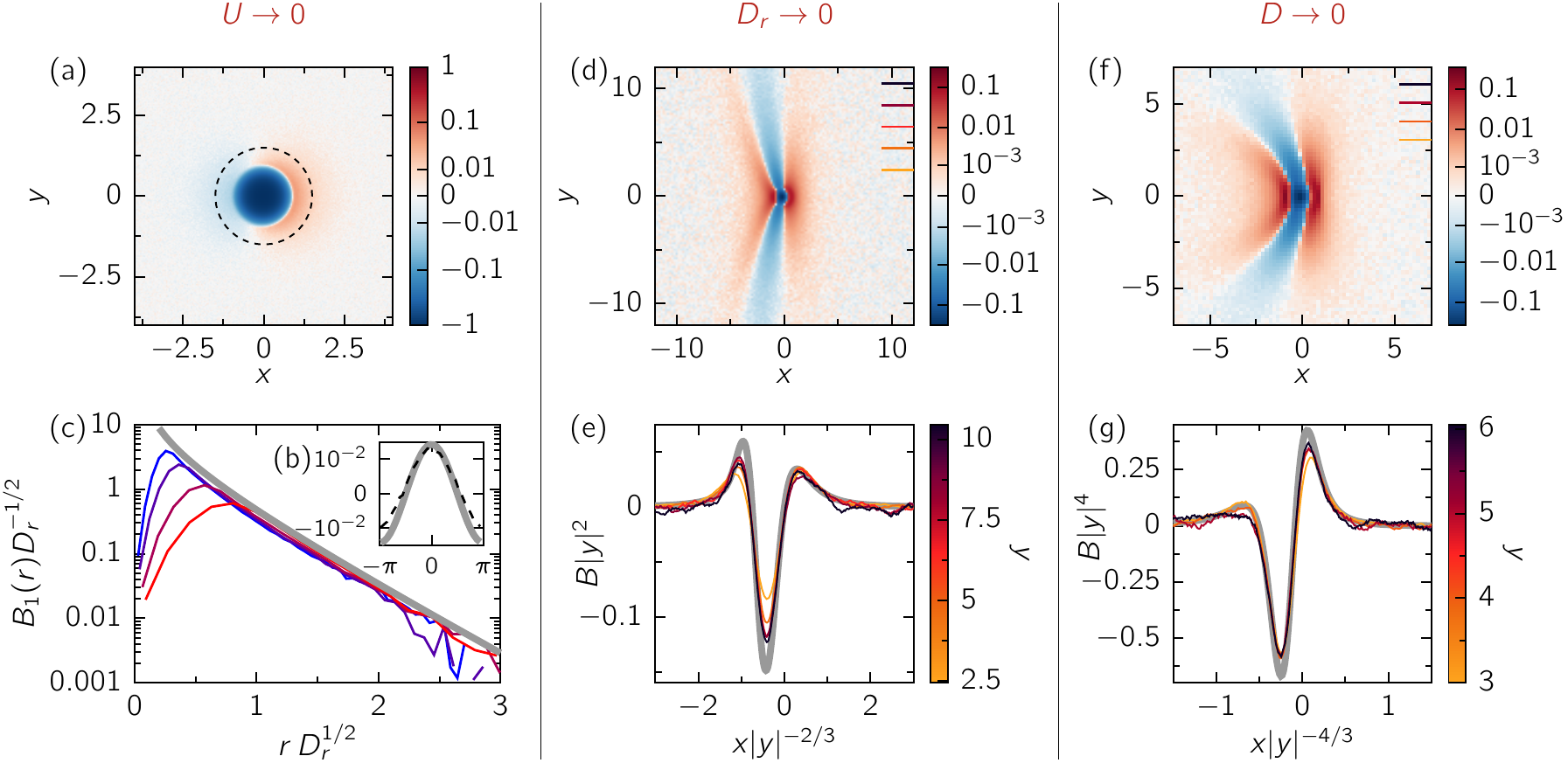}
\caption{
Limiting regimes.
(a)-(c) Small propulsion ($U\to 0$). 
    (a) Numerical correlation ($\phi = 0.04$ and $\epsilon\ratio D\ratio D_r\ratio U = 1\ratio 0.1\ratio 0.5\ratio 0.05 $).
    (b) Cut of the previous correlation at radius $r_0=1.5$ (dashed black line), and theoretical prediction
    (gray line).
    (c) Rescaled plot of the first Fourier coefficient $B_1(r)$ for $\phi = 0.04$ and $\epsilon\ratio D\ratio U = 1\ratio 0.1\ratio 0.05$ for $D_r = 0.1, 0.2, 0.5, 1$ (blue to red), the gray line is the theoretical prediction at large distance.
(d)-(e) No rotational diffusion ($D_r = 0$).
    (d) Numerical correlation
    ($\phi = 0.04$ and $\epsilon\ratio D\ratio D_r\ratio U =1\ratio 0.1\ratio 0\ratio 10$).
    (e) Rescaled horizontal cuts of the correlation from (d), the gray line is the theoretical prediction~\eqref{eq:B0_Dr0_int}.
(f)-(g) No translational diffusion ($D=0$). 
    (f) Numerical correlation
    ($\phi = 0.04$ and $\epsilon\ratio D\ratio D_r\ratio U =1\ratio 0\ratio 0.1\ratio 10$).
    (g) Rescaled horizontal cuts of the correlation from (f) according to Eq.~\eqref{eq:scaling_D0}.
}
\label{fig:small_U}
\end{figure*}

For small propulsion velocity $U$, which corresponds to the left side of the phase diagram, Eq.~(\ref{eq:correl_dir}) can be solved perturbatively as shown in App.~\ref{app:th_smallU}.
To order one in $U$, we get a dipolar correction to the equilibrium radial correlation $h_\mathrm{eq}(r)$: $B(r,\theta)=h_\mathrm{eq}(r)+U\cos(\theta)B_1(r)$, where the first Fourier coefficient $B_1(r)$ decays exponentially over a length $\ell_U$.
This prediction is compared to the numerical simulations in Fig.~\ref{fig:small_U}(a)-(c); a quantitative agreement with the prediction is obtained, without any adjustable parameter.
Note that at order $U^2$, an attractive correction to the equilibrium pair correlation $h_\mathrm{eq}(r)$ is found, 
which is compatible with the results obtained when the activity is introduced in the form of a colored noise~\cite{Farage2015, MariniBettoloMarconi2015, Fodor2016b}.

We turn to the large activity regime, and focus first on the limit $D_r=0$, where the wings are entirely deployed~(Fig.~\ref{fig:small_U}(d)); this limit corresponds to the bottom right corner of the phase diagram~(Fig.~\ref{fig:phase_diag}).
Equation (\ref{eq:correl_dir}) can be solved in Fourier space (App.~\ref{app:th_noDr}), yielding
\begin{equation}\label{eq:B0_Dr0_int}
\tilde B(\kk) = -\frac{k^2\tilde V(k)}{2\pi U}\int_0^{2\pi} \frac{d\theta'}{2\ell_r k^2+i\kk\cdot(\hat\ee_0-\hat\ee_{\theta'})}.
\end{equation}
The integrand has the form of the correlations in a driven binary mixture~\cite{Poncet2017}, and the integration over $\theta'$ mirrors that here there are particles moving in all directions.
At distances below $\ell_r$, Eq.~(\ref{eq:B0_Dr0_int}) yields a dipolar correlation, as in the small activity regime. 
Beyond $\ell_r$, the form of the correlations can be obtained by performing the integral and focusing on the singularity at the origin in Fourier space:
\begin{equation}
B(x,y) \underset{r\gg\ell_r}{\simeq} \frac{\tilde V(0)}{D_0} \frac{1}{y^2} F \left(\frac{x}{\ell_r^{1/3}|y|^{2/3}} \right),\label{eq:B0_Dr0_scaling}
\end{equation}
where the explicit form of the scaling function $F(u)$ is given in App.~\ref{app:th_noDr}.
The scaling (\ref{eq:B0_Dr0_scaling}) can be tested by plotting rescaled cuts of the correlation at different values of $y$ (Fig.~\ref{fig:small_U}(e)); as predicted, the cuts collapse to the scaling function $F(u)$.

We now address the shape of the wings in presence of a finite rotational diffusion $D_r$.
We take the limit $D=0$, which corresponds to the top right corner of the phase diagram~(Fig.~\ref{fig:phase_diag}), where the only available lengthscale is the persistence length $\ell_p$ (App.~\ref{app:th_noD}). 
Beyond $\ell_p$, particles loose the memory of their orientation and their correlation vanish.
Below $\ell_p$, the solution of Eq.~(\ref{eq:correl_dir}) is dominated by angles close to $(\theta,\theta')\simeq (0, 0)$ and $(\pi, \pi)$,
and we obtain the following scaling form: 
\begin{equation}\label{eq:scaling_D0}
B(x,y)\underset{r\ll\ell_p}{\simeq} \frac{\tilde V(0)}{D_r} \frac{\ell_p^4}{y^4}G \left(\frac{\ell_p^{1/3} x}{|y|^{4/3}} \right);
\end{equation}
the function $G$ can be obtained numerically.
This prediction is in very good agreement with the numerical simulations (Fig.~\ref{fig:small_U}(f),(g)).
Last, the transition with the scaling form obtained in the limit $D_r=0$ can be obtained by matching the widths of the two profiles, $x\sim \ell_r^{1/3}y^{2/3}$ (Eq.~(\ref{eq:B0_Dr0_scaling})) and $x\sim \ell_p^{-1/3}y^{4/3}$ (Eq.~(\ref{eq:scaling_D0})), leading to $y\sim\sqrt{\ell_r\ell_p}=\ell_U$.

The structure of the correlations is now characterized (Fig.~\ref{fig:phase_diag}).
For $\Pe<1$, the activity generates a dipolar correction to the equilibrium correlations for lengths $r<\ell_U$.
For $\Pe>1$, the dipolar correction crosses-over to depletion wings at a scale $\ell_r$; the shape of the wings is given by the scaling forms (\ref{eq:B0_Dr0_scaling}) for $\ell_r<r<\ell_U$ and (\ref{eq:scaling_D0}) for $\ell_U<r<\ell_p$; the wings decay exponentially beyond $\ell_p$.
The behavior of the correlations is summarized on the phase diagram, Fig.~\ref{fig:phase_diag}.

For weak interactions, we have obtained a quantitative agreement between our predictions derived from Eq.~(\ref{eq:correl_dir}) and the numerical simulations.
Moreover, the left hand side of Eq.~(\ref{eq:correl_dir}), which controls the structure of the correlations, is also present in the two-body Smoluchowski equation~\cite{Hartel2018}, which is valid in the dilute limit for any interaction strength.
We conclude that the structure of the correlations predicted here holds in the dilute limit for any interaction strength, which we check in App.~\ref{app:sim_hard} with numerical simulations of hard particles. 

We finally note that, in practice, $D$ and $D_r$ are not taken as independent parameters.
For spherical particles, where $D\sim a^2D_r$, the P\'eclet number defined here takes the more common forms $\Pe=aU/D=U/(aD_r)$, and $\ell_U\sim a$.
At small P\'eclet, the dipolar correction, which decays over a length $\ell_U$, should thus be barely observable.
At large P\'eclet, depletion wings with the shape (\ref{eq:scaling_D0}) should be observed for $a<r<\ell_p=a\Pe$.
Some theoretical studies assume $D=0$~\cite{Solon2015d,Basu2018,Caprini2020} and define the P\'eclet number as $\Pe'=U/(a D_r)$; here also depletion wings with the shape (\ref{eq:scaling_D0}) should be observed below $\ell_p$.

\section{Experiments}\label{}

To confront our findings to real systems, we perform experiments with Janus particles, whose dynamics is similar to that of ABP~\cite{Howse2007, Ginot2015, Mano2017}.
Our Janus particles are propelled by a vertical AC electric field and swim in a two-dimensional horizontal plane~\cite{Nishiguchi2015, Nishiguchi2018} (App.~\ref{app:exp}, Fig.~\ref{fig:experiments}(a) and Movie 1 in the Supplemental Material).
We estimate the  experimental values of $U$, $D$ and $D_r$ in App.~\ref{app:exp_params}. This allows us to place the experiments on the phase diagram (dark blue arrow in Fig.~\ref{fig:phase_diag}), with a P\'eclet number $\Pe\simeq 90$: depletion wings are expected.

\begin{figure}
\begin{center}
\includegraphics[scale=1]{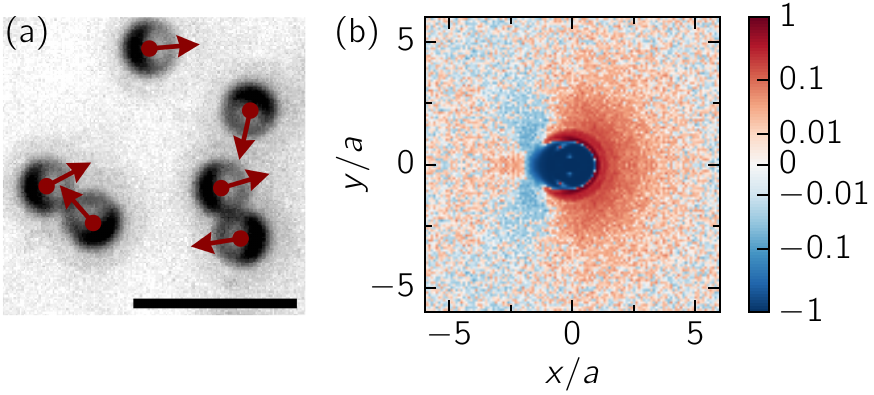}
\end{center}
\caption{Experiments
(a) Experimental image, indicating the detected positions and orientations of the Janus particles (red dots and arrows). 
Scale bar: $10~\si{\micro m}$.
(b) Pair correlation $B(\rr)$.}
\label{fig:experiments}
\end{figure}

The measured pair correlation is shown in Fig.~\ref{fig:experiments}(b) (additional figures for other values of the electric field are provided in App.~\ref{app:exp_Efield}).
Depletion wings are present, showing that they are a robust feature of pair correlations of self-propelled particles at high activity.
At short range, qualitative differences with the numerical simulations can be noted, such as depletion behind the active particle ($x<-a$, $y=0)$ before correlations turn positive.
These differences can be attributed to unavoidable hydrodynamic or electrostatic interactions~\cite{Zhang2020}.
Depletion wings are present at larger distances, but their decay cannot be characterized due to insufficient statistics.
Whether or not these wings follow the scaling laws predicted here or exhibit different exponents is an open question.

\section{Conclusion}\label{}

We have unveiled two regimes where active polar particles without alignment interactions have a pair correlation with a self-similar shape and an algebraic decay characterized by anomalous exponents.
In presence of velocity-orientation coupling~\cite{Iwasawa2020, Deseigne2010, Dauchot2019} or alignment interactions~\cite{Iwasawa2020, Bricard2013, Zhang2020}, we may expect distinct scaling laws to appear in the correlations.
As we have shown, the correlations have a rich structure even without three-body interactions, it would therefore be instructive to measure them in a dilute configuration first, and then to see how they evolve as density increases.

\begin{acknowledgments}
The authors would like to thank K. A. Takeuchi for his hospitality and useful discussions,  J. Iwasawa for his help with experimental and image processing protocols, and D. Bartolo, O. Dauchot and A. Solon for insightful discussions.
D.N. was supported by JSPS KAKENHI Grant Numbers JP19K23422, JP19H05800 and JP20K14426.
\end{acknowledgments}

\appendix

\begin{widetext}
\section{Numerical simulations} \label{app:sim}
\subsection{Description}\label{app:sim_desc}
We consider $N$ particles ($N = 5000$) in a square periodic box of size $L = \sqrt{\frac{N}{\rho}}$ with $\rho$ the density ($L\sim 300$ at $\rho = 0.05$). Initially the positions $\rr_i$ and the orientations $\theta_i$ of the particles
are assigned uniformly at random.

We use stochastic molecular dynamics and consider the following Langevin equations
\begin{align}
\dot \rr_i &= -\nabla_i \sum_{j\neq i} V(\rr_i-\rr_j)
+ U \hat \ee_{\theta_i} + \sqrt{2D} \eeta_i, \label{eq:X} \\
\dot\theta_i &= \sqrt{2D_r} \nu_i. \label{eq:theta}
\end{align}
$U$ is the velocity, $D$ the transitional diffusion and $D_r$ the rotational diffusion. The mobility $\gamma$ is set to $1$.
$ \eeta_i$ and $\nu_i$ are Gaussian white noises with unit variance.
During a time increment $\Delta t$ ($\Delta t = 0.05$ for the simulations presented in the article), the term $\sqrt{2D_r} \nu_i$ generates an
increment $\upsilon\sqrt{2D_r\Delta t}$ with $\upsilon$ a random number generated from a standard normal distribution. Similarly for $\sqrt{2D} \eeta_i$.

We use the following soft-sphere potential, where the diameter $a$ of a particle and the potential strength $\epsilon$ are set to unity,
\begin{equation}
	V(\rr) = \begin{cases}
		\frac{1}{2} (1-\|\rr\|)^2 & \mbox{if } \|\rr\| \leq 1 \\
		0 & \mbox{otherwise.}
	\end{cases}
\end{equation}

We let the system evolve for a time $t_\text{eq}  \sim 500$ before starting to record the correlations.
$B(\xx)$ is then recorded with a spatial resolution $\Delta x=0.1$, with one measure every $\Delta t_c = 5$, during an overall time period $T\sim 5\cdot 10^5$. The results are averaged over 100 realizations of the system.

\subsection{Simulations of dilute hard particles} \label{app:sim_hard}
The theoretical approach that we use is a linearized Dean equation, a framework which is valid for weak interactions. Consequently, the simulations that we compare to our theoretical results have relative values of potential strength and velocity $\epsilon\ratio U = 1\ratio 10$. This means that the particles are able to interpenetrate. In this weak interaction regime, we are able to test quantitatively the scalings that we obtain.

That being said, it is interesting to test the scaling exponents that we obtain on simulations in the regime of strong interactions. We do it in Fig~\ref{smfig:hard} and find that indeed the cuts of the correlation function obey the predicted scalings both in the limit of low rotational diffusion and in the limit of low translational diffusion. This leads us to state that the scaling exponents that we found are robust.
Note that the limit curves are different from our prediction in the weak interaction regime. 

\begin{figure}
	\begin{center}
		\includegraphics[scale=1]{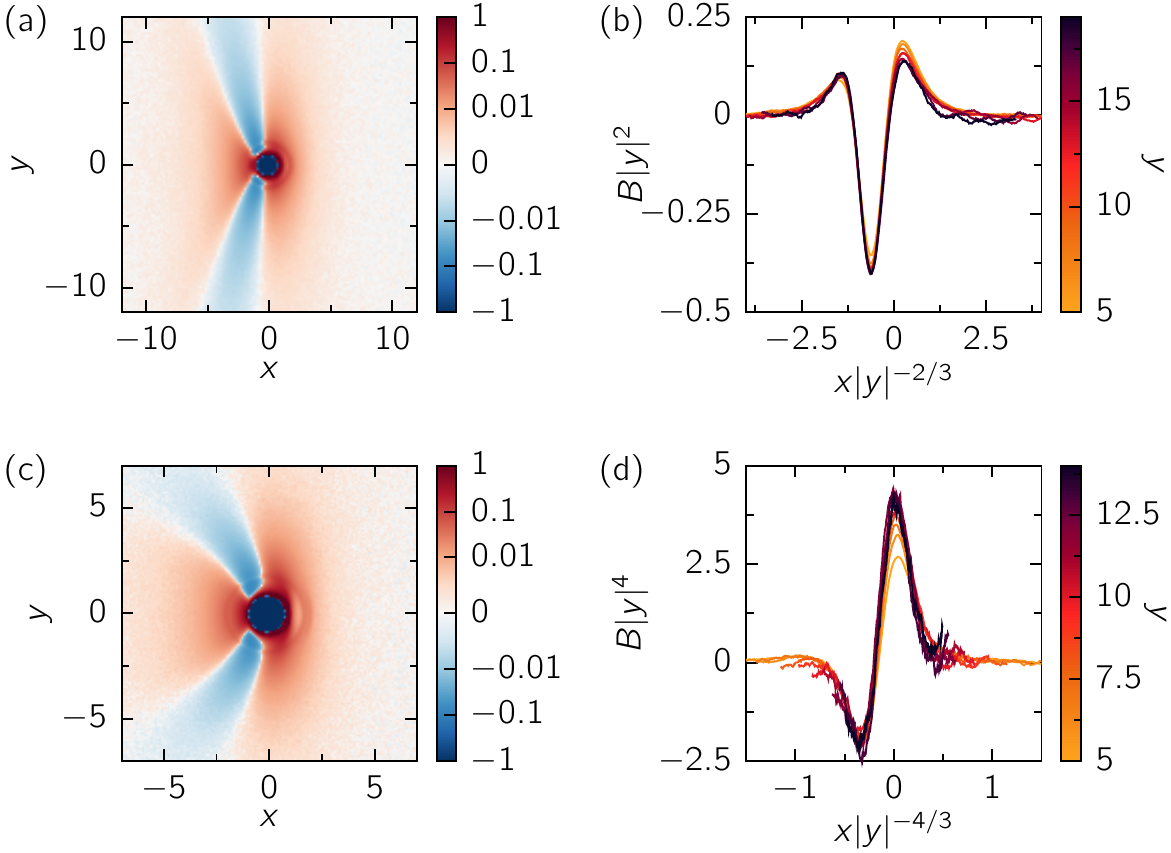}
	\end{center}
	\caption{Simulations in the limit cases $D_r = 0$ and $D = 0$ for dilute hard particles. (a) No rotational diffusion. Correlation $B(\rr)$ for $\rho = 0.05$, $\epsilon\ratio D\ratio D_r\ratio U = 50\ratio 0.1\ratio 0 \ratio 10$. (b) Rescaled cuts with the exponents predicted in the article. 
	(c) No translational diffusion. Correlation $B(\rr)$ for $\rho = 0.02$, $\epsilon\ratio D\ratio D_r\ratio U = 50\ratio 0 \ratio 0.1 \ratio 10$.
	(d) Rescaled cuts with the exponents predicted in the article. 
}
	\label{smfig:hard}
\end{figure}

\subsection{Extension of the dilute regime}\label{app:sim_extension}

Here we discuss the range of density $\rho$ where the dilute approximation that we use should be valid.
We focus on hard particles, and expect range of density to be wider for soft particles.
There are two ways to estimate where the dilute regime sits.
The first is to look at the low density branch of the motility induced phase separation (MIPS) line on the phase diagram for ABP~\cite{Digregorio2018}.
As MIPS is a many-body effect, the density should be far below this line to remain in the dilute regime. 
For instance, for $\Pe=100$, $\rho_\textrm{MIPS}\simeq 0.18$ ($\phi_\textrm{MIPS}\simeq 0.14$).

The second way is to look at measured pair correlations.
By definition, in the dilute regime the pair correlation does not depend on the density.
We plot the pair correlation for densities $\rho=0.01$, $0.02$ and $0.05$ for $D=0$ and $\Pe'=100$ on Fig.~\ref{smfig:dilute_hard}.
We observe an important difference between $\rho=0.02$ and $\rho=0.05$, while the difference between $\rho=0.01$ and $\rho=0.02$ is much smaller.
We conclude that the limit of the dilute regime for this P\'eclet number sits around $\rho\simeq 0.02$ ($\phi\simeq 0.016$). 


\begin{figure}
\begin{center}
\includegraphics[scale=1]{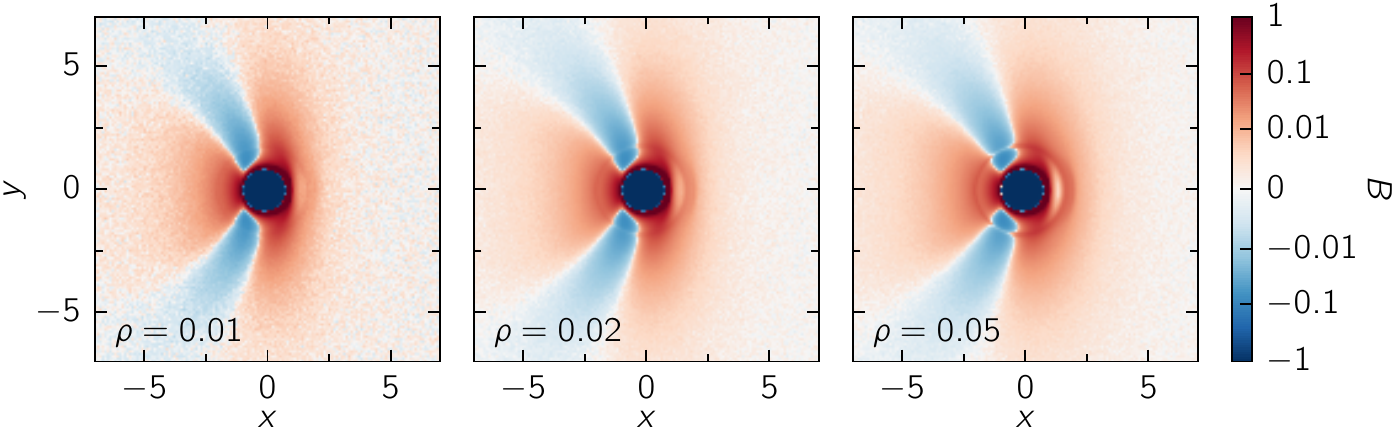}
\end{center}
\caption{Simulations in the limit case $D=0$ for hard particles, $\epsilon\ratio D\ratio D_r\ratio U = 50\ratio 0 \ratio 0.1 \ratio 10$, for densities $\rho=0.01$, $0.02$, $0.05$.}
\label{smfig:dilute_hard}
\end{figure}

\section{Theoretical computations} \label{app:th}

\subsection{Equation for the correlations} \label{app:th_eq}

\subsubsection{Dean equation for Active Brownian Particles}

The Langevin equations for Active Brownian Particles read,
\begin{align}
\dot \rr_i &= -\nabla_i \sum_{j\neq i} V(\rr_i-\rr_j) \label{smeq:X}
+ U \hat \ee_{\theta_i} + \sqrt{2D} \eeta_i  \\
\dot\theta_i &= \sqrt{2D_r} \nu_i \label{smeq:theta}
\end{align}
$D$ is the translational diffusivity, $D_r$ is the rotational diffusivity, $\eeta_i$ and $\nu_i$ are Gaussian white noises of unit variance. In this supplementary material, we set the particle diameter $a=1$ and the friction coefficient $\gamma = 1$.

We define the density in position-orientation space $f(\rr, \theta, t)$ as 
\begin{align}
f(\rr, \theta, t) &= 
\sum_{i=1}^N \sum_{m=-\infty}^\infty f_i(\rr, \theta + 2m\pi, t) \label{smeq:def_rho} \\
f_i(\rr, \theta, t) &= \delta(\rr_i(t) - \rr)
\delta(\theta_i(t) - \theta).
\end{align}

We consider a smooth and fastly decaying test function $\varphi(\rr, \theta)$. By definition of $f_i$,
\begin{equation}
 \varphi(\rr_i(t), \theta_i(t)) = \int d\rr \int_{-\infty}^\infty d\theta  f_i(\rr, \theta, t) \varphi(\rr, \theta).
\end{equation}
Then, the time derivative of $\varphi(\rr_i(t), \theta_i(t))$ can be written in two different ways:
\begin{equation}
 \frac{d}{dt}\varphi(\rr_i(t), \theta_i(t)) = \int d\rr \int_{-\infty}^\infty d\theta \frac{\partial f_i}{\partial t}(\rr, \theta, t) \varphi(\rr, \theta)
 = \int d\rr \int_{-\infty}^\infty d\theta f_i(\rr, \theta, t) (dt)^{-1} d\varphi(\rr, \theta),
\end{equation}
with the differential $d\varphi$ given by the It\^{o} formula~\cite{Oksendal_2013},
\begin{align}
 d\varphi &= \nabla \varphi \cdot d\rr_i + \frac{\partial \varphi}{\partial\theta}d\theta_i+ \frac{1}{2}\nabla^2 \varphi (d\rr_i)^2 + \frac{1}{2}\frac{\partial^2 \varphi}{\partial\theta^2} d\theta_i^2 + \frac{\partial}{\partial\theta}\nabla \varphi\cdot d\rr_i d\theta_i \\
 &= \nabla \varphi \cdot\left\{-\nabla_i \sum_j V(\rr_i-\rr_j)
 + U \hat \ee_{\theta_i}\right\} dt
 + D dt \nabla^2 \varphi + D_r dt \frac{\partial^2 \varphi}{\partial\theta^2}.
\end{align}
The differentials are computed from Eqs.~\eqref{smeq:X} and \eqref{smeq:theta} (we assume $\nabla V(\vec 0) = \vec 0$). Performing integrations by part and recalling that $\varphi$ is arbitrary, one obtains
\begin{equation}
\frac{\partial f_i}{\partial t}
= D \nabla^2 f_i+ D_r\frac{\partial^2 f_i}{\partial^2\theta}
+ \nabla\left(f_i \sum_{j=1}^N \nabla V(\rr - \rr_j(t))\right)
- U\ee_\theta \cdot \nabla f_i
- \sqrt{2D} \nabla f_i\cdot \eeta_i
- \sqrt{2D_r} \frac{\partial}{\partial\theta} (f_i\nu_i).
\label{smeq:dean}
\end{equation}

Using Eq.~\eqref{smeq:def_rho} and rearranging the noises like Dean~\cite{Dean1996}, we finally obtain the following Dean equation for $f(\rr, \theta, t)$,
\begin{equation}
\label{smeq:Dean}
\frac{\partial}{\partial t} f(\rr, \theta, t) = -\nabla \vec J(\rr, \theta, t)
- \frac{\partial}{\partial\theta} K(\rr, \theta, t)
\end{equation}
with the currents
\begin{align}
\vec J(\rr, \theta, t) &= -D \nabla f(\rr, \theta, t)
- f(\rr, \theta, t) \int_0^{2\pi} d\theta (\nabla V \ast f)(\rr, \theta, t)
+ f(\rr, \theta, t) U \hat \ee_\theta
- f^{1/2}(\rr, \theta, t)
\eeta(\rr, \theta, t) \\
K(\rr, \theta, t) &= -D_r\frac{\partial}{\partial\theta} f(\rr, \theta, t)
-f^{1/2}(\rr, \theta, t) \nu(\rr, \theta, t).
\end{align}
The spatial convolution is defined by $(f\ast g)(\rr) = \int d\rr' f(\rr')g(\rr-\rr')$.
$\eeta$ and $\nu$ are Gaussian white noises with correlations
\begin{align}
\langle \eta^\alpha(\rr, \theta, t) \eta^\beta(\rr, \theta, t) \rangle
= 2 D \delta^{\alpha\beta} \delta(\rr-\rr') \delta(\theta-\theta') \delta(t-t'),\\
\langle \nu(\rr, \theta, t) \nu(\rr, \theta, t) \rangle
= 2 D_r \delta(\rr-\rr') \delta(\theta-\theta') \delta(t-t').
\end{align}

\subsubsection{Linearized Dean equation}

The Dean equation for Active Brownian Particles is nonlinear with multiplicative noise.
It is thus very difficult to tackle. Our approximation consists in linearizing it
around an homogeneous density $\rho$ which is the average density of particles.
We write
\begin{equation}
f(\rr, \theta, t) = \frac{\rho}{2\pi} + \sqrt\frac{\rho}{2\pi} \phi(\rr, \theta, t).
\end{equation}
The field $\phi$ is assumed to be of order $1$ in $\rho$.

At the lowest order, the Dean equation \eqref{smeq:Dean} becomes linear with additive noise.
\begin{equation}
\frac{\partial\phi}{\partial t} = \left[ D \nabla^2
+ D_r\frac{\partial^2}{\partial\theta^2}
- U \hat\ee_\theta \cdot\nabla \right] \phi
+ \frac{\rho}{2\pi}\int_0^{2\pi} d\theta' (\nabla^2 V\ast\phi)(\theta')
+ \nabla\cdot\eeta + \frac{\partial \nu}{\partial\theta}.
\label{smeq:linearDean}
\end{equation}


\subsubsection{Equation for the correlations}

We define the following correlations
\begin{align}
\mcC(\rr_1, \rr_2, \theta_1, \theta_2)
&=  \langle \phi(\rr_1, \theta_1) \phi(\rr_2, \theta_2)\rangle, \\
C(\rr, \theta, \theta') &= \frac{1}{\rho} \left\{\mcC(\vec 0, \rr, \theta, \theta') -  \delta(\rr)\delta(\theta-\theta') \right\}.
\end{align}
One checks that the definition of $C(\rr, \theta, \theta')$ is consistent with Eq.~(1) the article.
We use It\^{o} calculus to compute the time evolution of $\mcC$.
\begin{multline}
\mcC(\rr_1, \rr_2, \theta_1, \theta_2, t+\delta t) - \mcC(\rr_1, \rr_2, \theta_1, \theta_2, t) = \\
\langle \phi(\rr_1, \theta_1, t) \delta \phi(\rr_2, \theta_2, t) \rangle
+ \langle \delta\phi(\rr_1, \theta_1, t)  \phi(\rr_2, \theta_2, t) \rangle
+ \langle \delta \phi(\rr_1, \theta_1, t) \delta \phi(\rr_2, \theta_2, t)\rangle.
\end{multline}

Computing the terms from the linearized Dean equation~\eqref{smeq:linearDean}, one shows that
\begin{multline}
\partial_t \mcC(\rr_1, \rr_2, \theta_1, \theta_2)
= \left[ D(\nabla_1^2 + \nabla_2^2)
+ D_r(\partial_{\theta_1}^2 + \partial_{\theta_2}^2)
- U (\hat\ee_{\theta_1} \cdot\nabla_1 + \hat\ee_{\theta_2} \cdot\nabla_2) \right] \mcC(\rr_1, \rr_2, \theta_1, \theta_2) \\
+ \frac{\rho}{2\pi}\int_0^{2\pi} d\theta' \left[
\nabla_1^2 V\ast\mcC(\rr_1, \rr_2, \theta', \theta_2)
+ \nabla_2^2 V\ast\mcC(\rr_1, \rr_2, \theta_1, \theta')
\right]
+ \left[2D\nabla_1\nabla_2 + 2D_r\partial_{\theta_1}\partial_{\theta_2}\right]\delta(\rr_1-\rr_2)\delta(\theta_1 - \theta_2).
\end{multline}

We use the invariance of the system by translation and write the equation in terms of $C(\rr, \theta, \theta')$,
\begin{multline}
\partial_t C(\rr, \theta, \theta') =
 \left[ 2D \nabla^2 
+ D_r(\partial_{\theta}^2 + \partial_{\theta'}^2)
+ U (\hat\ee_{\theta} - \hat\ee_{\theta'})\cdot\nabla \right] C(\rr, \theta, \theta')
+ 2\nabla^2 V(\rr) \\ +
\frac{\rho}{2\pi}\int_0^{2\pi} d\theta'' \nabla^2 V \ast\left[C(\rr, \theta, \theta'')
+ C(\rr, \theta'', \theta')\right].
\end{multline}

The conventions are such that the pair correlation function $C(\rr, \theta, \theta')$ is normalized with respect to the density $\rho$. Focusing on the low density regime ($\rho\to 0$), we can neglect the convolution of $C$ with the potential $V$.
The equation that we consider is 
\begin{equation} \label{smeq:eq_C0_time}
\partial_t C(\rr, \theta, \theta')
\underset{\rho\to 0}{=}
\left[ 2D \nabla^2 
+ D_r(\partial_{\theta}^2 + \partial_{\theta'}^2)
+ U (\hat\ee_{\theta} - \hat\ee_{\theta'})\cdot\nabla \right] C(\rr, \theta, \theta')
+ 2\nabla^2 V(\rr).
\end{equation}
Note that for a passive system, the solution of this equation is the direct correlation function (the one involved in the Ornstein-Zernike equation).
In the following, we are only interested in the stationary correlations which satisfy the following linear partial differential equation.
\begin{gather}
\left[ 2D \nabla^2 
+ D_r(\partial_{\theta} + \partial_{\theta'})
+ U (\hat\ee_{\theta} - \hat\ee_{\theta'})\cdot\nabla \right] C(\rr, \theta, \theta')
= -2\nabla^2 V(\rr).
\label{smeq:eq_C0}
\end{gather}

By rotational invariance, $C(\rr,\theta,\theta')=C(\mcR_{-\theta}\cdot\rr,0,\theta'-\theta)$. Like in the article, we define the ``profile seen by a particle in its reference frame'',
\begin{align}
B(\rr) &= \frac{1}{2\pi}\int_0^{2\pi} C(\rr, 0, \theta') d\theta'.
\end{align}

\subsection{Small activity} \label{app:th_smallU}
At small activity, $U\ll 1$, one can expand $C$ in power of $U$.
\begin{equation}
 C(\xx, \theta, \theta') = C^{(0)}(\xx) + U C^{(1)}(\xx, \theta, \theta') + U^2 C^{(2)}(\xx, \theta, \theta') + \dots
\end{equation}

Let us write Eq~\eqref{smeq:eq_C0} in Fourier space, using the convention $\tilde C(\kk, \theta, \theta', t) = \int d\rr e^{-i\kk\rr} C(\rr, \theta, \theta', t)$.
\begin{equation}
\left[ -2D k^2
+ D_r(\partial_{\theta} + \partial_{\theta'})\right] \tilde C(\kk, \theta, \theta')
=  2k^2 \tilde V
-iU \kk\cdot (\hat\ee_{\theta} - \hat\ee_{\theta'}) \tilde C(\kk, \theta, \theta').
\end{equation} 

The passive correlation ($U = 0$) doesn't depend on the angles, the order $0$ of the
equation above leads to
\begin{equation}
\tilde C^{(0)}(\kk) = \frac{-\tilde V(\kk)}{D}.
\end{equation}
This is the usual Random Phase Approximation solution for direct correlations.
At order $1$, the equation to solve and its solution are
\begin{gather}
 \left[ -2D k^2
 + D_r(\partial_{\theta}^2 + \partial_{\theta'}^2)\right] \tilde C^{(1)}(\kk, \theta, \theta')
 = -i\kk\cdot (\hat\ee_{\theta} - \hat\ee_{\theta'}) \tilde C^{(0)}(\kk),  \\
 \tilde C^{(1)}(\kk, \theta, \theta') = i\kk\cdot (\hat\ee_{\theta} - \hat\ee_{\theta'}) \tilde D^{(1)}(\kk), \\
\tilde D^{(1)}(\kk) = \frac{\tilde C^{(0)}(\kk)}{2Dk^2 + D_r}.
\end{gather}
Next, we solve the second order,
\begin{gather}
\left[ -2D k^2
+ D_r(\partial_{\theta}^2 + \partial_{\theta'}^2)\right] \tilde C^{(2)}(\kk, \theta, \theta')
= 2k^2 (1 - \hat\ee_{\theta} \cdot \hat\ee_{\theta'}) \tilde D^{(1)}(\kk), \\
\tilde C^{(2)}(\kk, \theta, \theta') = \frac{-\tilde D^{(1)}(\kk)}{D} + \frac{k^2 \tilde D^{(1)}(\kk)}{Dk^2 + D_r} (\hat\ee_{\theta} \cdot \hat\ee_{\theta'}).
\end{gather}

At the end of the day, the expansions at order $U^2$ of
$\tilde C(\kk, \theta, \theta')$ and $\tilde B(\kk)$ are
\begin{align}
\tilde C(\kk, \theta, \theta') &= \frac{\tilde V(\kk)}{D}\left\{
\left(-1 + \frac{U^2}{D^2}\frac{1}{2k^2 + \ell_U^{-2}}\right)
- \frac{U}{D} \frac{i\kk\cdot (\hat\ee_{\theta} - \hat\ee_{\theta'})}{2k^2 + \ell_U^{-2}}
+ \frac{U^2}{D^2} (\hat\ee_{\theta} \cdot \hat\ee_{\theta'})\left(
\frac{1}{2k^2 + \ell_U^{-2}} - \frac{1}{k^2 + \ell_U^{-2}}
\right)
\right\} \\
\tilde B(\kk) &= \frac{\tilde V(\kk)}{D}\left\{
\left(-1 + \frac{U^2}{D^2}\frac{1}{2k^2 + \ell_U^{-2}}\right)
- \frac{U}{2\pi D} \frac{ik_x}{2k^2 + \ell_U^{-2}}
\right\} 
\end{align}
with the typical length scale $\ell_U = \sqrt{D_r/D}$.

Our goal is now to look at large distances, that is to say, small wave number $\kk$. We assume that $\tilde V(\kk)$ is regular at $0$ (short-range potential) and we make the substitution $\tilde V(\kk) \mapsto \tilde V(0)$. We define the function $\tilde G(\kk)$ and its inverse Fourier transform $G(\rr)$ as
\begin{equation}
\tilde G(\vec k) = \frac{1}{2\vec k^2 + \ell_U^{-2}} 
\qquad\Leftrightarrow\qquad
G(\rr) = \frac{1}{4\pi}K_{0}\left(\frac{\|\rr\|}{\sqrt{2} \ell_U}\right),
\end{equation}
with $K_0$ the modified Bessel function of the second kind of order $0$. In real space, the expansion of $B(\rr)$ becomes,
\begin{align}
B(\rr) &\equi{r\to\infty} - \frac{\tilde V(0)}{D} + 
U^2 \frac{\tilde V(0)}{D^3} G(\rr) - U \frac{\tilde V(0)}{2\pi D^2} \frac{\partial G(\rr)}{\partial x} \\
B(r, \theta) &\equi{r\to\infty} - \frac{\tilde V(0)}{D} + 
U^2 \frac{\tilde V(0)}{4\pi D^3} K_{0}\left(\frac{r}{\sqrt{2} \ell_U}\right) + U  \frac{\tilde V(0) \sqrt{D_r}}{2\pi \sqrt{2} D^{3/2}} K_{1}\left(\frac{r}{\sqrt{2} \ell_U}\right) \cos\theta
\label{smeq:dvtB}
\end{align}
$K_1$ is the modified Bessel function of the second kind of order $1$. We detail the meaning of the terms. The first one in the passive correlation. The second term, scaling as $U^2$ is a \textit{positive} contribution to the isotropic part of the correlations. As stated in the article, this has been argued, in the literature,  to account for the motility induced phase separation. The last term is the dipolar correlation, at order $U$, on which we focused in the article. Eq.~\eqref{smeq:dvtB} corresponds to the prediction in Fig.~3b-c. At large distance, the Bessel functions decay exponentially,
\begin{equation}
K_0(r) \equi{r\to\infty} K_1(r) \equi{r\to\infty} \sqrt\frac{\pi}{2r} e^{-r}.
\end{equation}
Both the dipolar contribution ($\cos\theta$) and the additional isotropic part thus decay exponentially over the length scale $\ell_U$.


\subsection{No rotational diffusion} \label{app:th_noDr}

We now focus on the limit of no rotational diffusion $D_r = 0$. We easily obtain the Fourier transform $C(\rr, \theta, \theta')$ from  Eq.~\eqref{smeq:eq_C0},
\begin{gather}
	\left[ 2D \nabla^2 
	+ U (\hat\ee_{\theta} - \hat\ee_{\theta'})\cdot\nabla \right] C(\rr, \theta, \theta')
	= -2\nabla^2 V(\rr), \\
	\tilde C(\kk, \theta, \theta') =
	\frac{-2k^2\tilde V(\kk)}{2D k^2 - iU\kk\cdot(\hat \ee_\theta - \hat \ee_{\theta'})}.
\end{gather}
This solution is very similar to the one for a binary mixture of particles forced respectively by $U\hat \ee_\theta$  and $U \hat\ee_{\theta'}$~\cite{Poncet2017}.
We integrate over $\theta'$ to obtain $B$, this gives,
\begin{equation}
\tilde B(\kk) = -\frac{k^2\tilde V(\kk)}{\pi D} \int_0^{2\pi}
\frac{d\theta'}{2k^2 - i\ell_r^{-1}\kk\cdot(\hat\ee_x - \hat\ee_{\theta'})}
= \frac{-2k^2\tilde V(\kk)}{D \sqrt{\left(2k^2 - i\ell_r^{-1} k_x\right)^2 + \ell_r^{-2}k^2}}
\end{equation}
with the characteristic length scale $\ell_r = D/U$. $k_x = \kk\cdot \hat\ee_x$, $k_y = \kk\cdot \hat\ee_y$ where $\hat\ee_x$ and $\hat\ee_y$ are the unit vector along the horizontal and vertical axes of the plane.

We consider the limit of large distance, that is to say $\kk\to\vec 0$. We assume that $\tilde V(\kk)$ is regular so that we can replace it by $\tilde V(0)$. Furthermore, an analysis of leading
terms gives $k_y^2 \sim k_x^3$. We keep up these leading terms:
\begin{equation}
 \tilde B(\kk) \equi{k\to 0} \frac{-2\tilde V(0) k_x^2}{D \sqrt{\ell_r^{-2} k_y^2-4i\ell_r^{-1} k_x^3}}.
\end{equation}
We can inverse Fourier transform with respect to $k_y$:
\begin{equation}
B(k_x, y) = -\frac{2 \tilde V(0) \ell_r}{\pi D} k_x^2 K_0\left(|y|\sqrt{-4i\ell_r k_x^3}\right).
\end{equation}
We then Fourier transform with respect to $k_x$: $B(x, y) = (2\pi)^{-1} \int dk_x e^{ik_x x} B(k_x, y)$. We perform the changes of variables $q = -(\ell_r y^2)^{1/3}k_x$, $w = x u^{1/3} (\ell_r y^2)^{-1/3}$ and obtain
\begin{align}
B(x,y) & \sim \frac{\tilde V(0)}{D} \frac{1}{y^2} F \left(\frac{x}{\ell_r^{1/3}|y|^{2/3}} \right),\\
F(w) & = \frac{1}{\pi^2}\int_{-\infty}^\infty e^{-iqw}q^2 K_0 \left(2\sqrt{iq^3} \right)dq.
\end{align}
This is the scaling form mentioned in the article, and used in Fig.~3e.

\subsection{No translational diffusion}  \label{app:th_noD}
We set $D=0$. Looking at distances large compared to the particle diameter, we replace $\tilde V(\kk)$ by $\tilde V(0)$. The equation we consider is 
\begin{equation} \label{eq:abp_Cd_D0}
\left[(\partial_{\theta}^2 + \partial_{\theta'}^2)
+ i \ell_p \kk\cdot (\hat\ee_{\theta} - \hat\ee_{\theta'}) \right] \tilde C(\kk, \theta, \theta')
= \frac{2k^2 \tilde V(0)}{D_r}.
\end{equation}
with the persistence length $\ell_p = U/D_r$.

At small $\ell_p\kk$ (large distance compared to $\ell_p$), the following development can be obtained,
\begin{equation}
\tilde C(\kk, \theta, \theta') = \frac{-\tilde V (0)}{D_r}\left\{
1 + i\ell_p \kk\cdot (\hat\ee_{\theta} - \hat\ee_{\theta'}) + \ell_p^2 k^2 \hat\ee_{\theta} \cdot \hat\ee_{\theta'} + \mathcal{O}\left((\ell_p k)^3\right) \right\}.
\end{equation}
This is a hint that $\tilde C(\kk, \theta, \theta')$ is analytical around $\kk = 0$, meaning that $C(\rr, \theta, \theta')$ decays fastly (e.g. exponentially) at distances large compared to $\ell_p$.

We now consider distances below $\ell_p$ but still large compared to the particle diameter.
We define the angles $\gamma$ and $\gamma'$ in the reference frame of $\kk$: $\kk\cdot \hat\ee_{\theta} = k\cos\gamma$ and $\kk\cdot \hat\ee_{\theta'} = k\cos\gamma'$, where $k$ is the norm of $\kk$.  We obtain
\begin{equation} \label{smeq:numeq_D0}
\left[(\partial_{\gamma}^2 + \partial_{\gamma'}^2)
+ i \ell_p k(\cos\gamma - \cos\gamma') \right] \tilde C(k, \gamma, \gamma')
= \frac{2k^2 \tilde V(0)}{D_r},
\end{equation}
and we study it in the regime $\ell_p k \gg 1$. A numerical resolution at constant $k$ (see Fig.~\ref{smfig:sol_D0}) shows that $\tilde C(k, \gamma, \gamma')$ concentrates around the two points $(\gamma, \gamma') = (0, 0)$ and $(\pi, \pi)$. We focus on $(0, 0)$ around which the equation reads
\begin{equation} \label{smeq:eq_D0}
\left[(\partial_{\gamma}^2 + \partial_{\gamma'}^2)
- \frac{i}{2} \ell_p k(\gamma^2 - \gamma'^2) \right] \tilde C(k, \gamma, \gamma')
= \frac{2\tilde V(0)}{D_r} k^2.
\end{equation}
We realize that we can inject the following scalings
\begin{equation}
\tilde C(k, \gamma, \gamma') \equi{\ell_p k \gg 1} \frac{2\tilde V(0)}{D_r} (\ell_p k)^{3/2} \tilde H\left(\gamma (\ell_p k)^{1/4}, \gamma' (\ell_p k)^{1/4}\right).
\end{equation}
From Eq.~\eqref{smeq:eq_D0}, the function $\tilde H(u, v)$, for $u$ and $v$ unbounded, is independent of $k$ and is the solution of the linear partial differential equation
\begin{equation} \label{eq:abp_D0_H}
\left[\partial_u^2 + \partial_v^2 - \frac{i}{2}\left(u^2 - v^2\right)\right] \tilde H(u, v) = 1.
\end{equation}
Around $(\gamma, \gamma') = (\pi, \pi)$, the scalings are
\begin{equation}
\tilde C(k, \gamma, \gamma') \equi{\ell_p k \gg 1} \frac{2\tilde V(0)}{D_r} (\ell_p k)^{3/2} \tilde H^\ast\left((\gamma-\pi) (\ell_p k)^{1/4}, (\gamma'-\pi) (\ell_p k)^{1/4}\right).
\end{equation}
with $H^\ast$ the complex conjugate of $H$.

The scaling for $\tilde B$ around $\gamma = 0$ is
\begin{align} \label{eq:abp_D0_Bd1}
\tilde B(k, \gamma) &= \frac{1}{2\pi}\int_0^{2\pi} d\gamma' \tilde C(k, \gamma, \gamma') 
\equi{\ell_p k \gg 1} \frac{\tilde V(0)}{\pi D_r} (\ell_p k)^{5/4} \tilde H_B\left(\gamma (\ell_p k)^{1/4}\right), \\
H_B(u) &= \int_{-\infty}^\infty dv\, H(u, v).
\end{align}
And around $\gamma = \pi$, one checks that 
\begin{align} \label{eq:abp_D0_Bd2}
\tilde B(k, \pi-\gamma) &
\equi{\ell_p k \gg 1} \frac{\tilde V(0)}{\pi D_r} (\ell_p k)^{5/4} \tilde H_B^\ast\left((\pi - \gamma) (\ell_p k)^{1/4}\right)
\end{align}
with $H_B^\ast$ the complex conjugate of $H_B$. We now switch from polar  coordinates $(k, \gamma)$ to cartesian coordinates $(k_x, k_y)$.
We approximate
\begin{align}
k_x &= k\cos\gamma \simeq \begin{cases}
+k & \mbox{if } \gamma\simeq 0 \\ -k & \mbox{if } \gamma\simeq \pi
\end{cases}, &
k_y &= k\sin\gamma \simeq \begin{cases}
k\gamma & \mbox{if } \gamma\simeq 0 \\ k(\pi-\gamma) & \mbox{if } \gamma\simeq \pi
\end{cases}.
\end{align}
As we consider small angles ($|\gamma|\ll 1$ or $|\pi-\gamma|\ll 1$), the values of $k_x$ and $k_y$ are such that $k_y\ll k_x$.

 Using the two expressions Eqs~\eqref{eq:abp_D0_Bd1} and \eqref{eq:abp_D0_Bd2}, we obtain a scaling form for the Fourier transform $\tilde B$.
\begin{equation}
\tilde B\left(k_x \gtrless 0, k_y\right) = \frac{\tilde V(0)}{\pi D_r} (\ell_p |k_x|)^{5/4} \tilde H_B^\pm \left(\frac{\ell_p^{1/4} k_y}{|k_x|^{3/4}}\right)
\end{equation}
with $\tilde H_B^+ = \tilde H_B$ used when $k_x>0$ and $\tilde H_B^- = \tilde H_B^\ast$ used when $k_x < 0$.
Finally, we perform the Fourier inversion
\begin{equation}
B(x, y) = \frac{1}{2\pi} \int dk_x e^{ik_x x} \frac{1}{2\pi} \int dk_y e^{ik_y y} \tilde B\left(k_x, k_y\right).
\end{equation}
first with respect to $k_y$, then with respect to $k_x$. Using the appropriate changes of variables, we obtain a scaling form for $B(x, y)$,
\begin{align} \label{eq:abp_D0_scale1}
B(x, y) &= \frac{\tilde V(0)}{\pi D_r} \frac{\ell_p^4}{y^4} G\left(\frac{\ell_p^{1/3} x}{|y|^{4/3}}\right), \\ \label{eq:abp_D0_scale2}
G(w) &= \frac{1}{2\pi}
\int_0^\infty dz e^{iwz} z^2 H_B^+(z^{3/4})
+ \frac{1}{2\pi}
\int_{-\infty}^0 dz e^{iwz} z^2 H_B^-(|z|^{3/4})
\end{align}
$H_B^\pm$ is the inverse Fourier transform of $\tilde H_B^\pm$, one has $H_B^-(a) = (H_B^+)^\ast(-a)$.
This form corresponds to Eq.~(10) of the article. The rescaled cuts are plotted on Fig.~3g with the prediction (gray line) computed from a rescaling of the numerical solution of Eq.~(7).

\begin{figure}
	\begin{center}
		\includegraphics[scale=1]{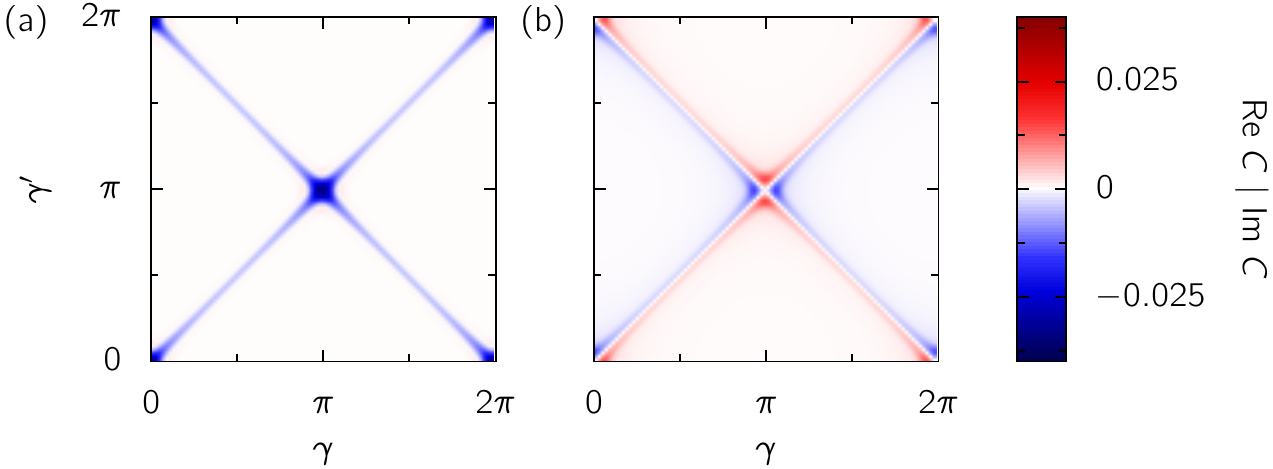}
	\end{center}
	\caption{Numerical resolution of Eq.~\eqref{smeq:numeq_D0} with $\tilde V(0) = U = D_r = 1$, for $k = 10^4$. (a) Real part of $C$. (b) Imaginary part of $C$. The solution concentrates around $(\gamma, \gamma') = (0, 0)$ and $(\pi, \pi)$. The numerical resolution consists in discretizing $\gamma$ and $\gamma'$, and then solving the linear system corresponding to Eq.~\eqref{smeq:numeq_D0}.}
	\label{smfig:sol_D0}
\end{figure}

\section{Numerical integration of the theoretical equation}  \label{app:numint}
\subsection{Equation in terms of three parameters}
We consider the time-dependent equation~\eqref{smeq:eq_C0_time} for the correlations $C$ at low density,
\begin{equation}
\partial_t C(\rr, \theta, \theta') =
\left[ 2D \nabla^2 
+ D_r(\partial_{\theta}^2 + \partial_{\theta'}^2)
+ U (\hat\ee_{\theta} - \hat\ee_{\theta'})\cdot\nabla \right] C(\rr, \theta, \theta')
+ 2\nabla^2 V(\rr).
\end{equation}
In polar coordinates, we write $\rr = r\ee_\phi$. The later equation depends on four coordinates (plus time): $(r, \phi, \theta, \theta')$. By performing a rotation of angle
$\theta$, the symmetries allow to reduce the problem to three parameters $(r, \alpha, \beta)$,
\begin{align}
\alpha &= \phi - \theta & \beta = \theta' - \theta.
\end{align}
We write $\xx = r\ee_\alpha = (x, y)$, $C$ is then a function only of $\xx$ and $\beta$.
Its time evolution is given by
\begin{align}
\partial_t C(\xx, \beta) &=
\left[ 2D \nabla^2 
+ D_r \mathcal{L}_\text{angles}
+ U \left((1-\cos\beta) \frac{\partial}{\partial x} -\sin\beta \frac{\partial}{\partial y} \right)
\right] C(\xx, \beta)
+ 2\nabla^2 V(\xx), \label{smeq:eq_c0_int}\\
\nabla^2 &=  \frac{\partial^2}{\partial x^2} + \frac{\partial^2}{\partial y^2} \\
\mathcal{L}_\text{angles} &= \left(y^2 \frac{\partial^2}{\partial x^2} +
x^2 \frac{\partial^2}{\partial y^2}
-2 xy \frac{\partial^2}{\partial x\partial y} -x\frac{\partial}{\partial x} - y\frac{\partial}{\partial y}\right)
+ 2 \left(-y \frac{\partial}{\partial x} + x \frac{\partial}{\partial y}\right)
\frac{\partial}{\partial\beta} + 2\frac{\partial^2}{\partial\beta^2}
\end{align}

It is important to note that $B$ is given by the integration over $\beta$,
\begin{equation}
B(\xx) = \int_0^{2\pi} d\beta\, C(\xx, \beta).
\end{equation}

\subsection{Numerical integration}
We consider the domain $(x, y, \beta) \in
[-x_\mathrm{max}, x_\mathrm{max}] \times [0, y_\mathrm{max}] \times [-\pi, \pi]$.
We discretize it with steps $\Delta x$ in $x$, $\Delta y=\Delta x$ in $y$ and $\Delta\beta$ in $\beta$.

We start from $C(x, y, \beta, t=0) = 0$ and integrate Eq.~\eqref{smeq:eq_c0_int}
in time using an explicit Euler scheme with time step $\Delta t$. The differential operators are evaluted using
finite differences valid at order $(\Delta x)^2$ and $(\Delta \beta)^2$,
\begin{align}
 \frac{\partial}{\partial x}C(x, y, \beta) &= \frac{C(x+\Delta x) - C(x-\Delta x)}{2\Delta x}, &
 \frac{\partial^2}{\partial x^2}C(x, y, \beta) &= \frac{C(x+\Delta x) + C(x-\Delta x) - 2 C(x)}{(\Delta x)^2},
\end{align}
and so on.

The potential is
\begin{equation}
V(\xx) = \begin{cases}
\frac{1}{2} (1-\|\xx\|)^2 & \mbox{if } \|\xx\| \leq 1 \\
0 & \mbox{otherwise}
\end{cases}
\end{equation}
and its Laplacian is evaluated on the grid.

The boundary conditions are as follow:
\begin{itemize}
	\item Periodic boundary conditions for $\beta$;
	\item $C(-x_\mathrm{max}, y, \beta) = C(+x_\mathrm{max}, y, \beta) = C(x, y_\mathrm{max}, \beta) = 0$;
	\item We use the symmetry relation $C(x, -y, -\beta) = C(x, y, \beta)$ to
	impose the additional points $C(x, -\Delta y, \beta) = C(x, \Delta y, -\beta)$.
\end{itemize}

The numerical integration in time converges to the stationary solution at sufficiently large time. When the increment on $C$
over a time step $\Delta t$ is small enough, we output the stationary solution $C^\text{eq}(\xx, \beta)$ and its integral over $\beta$, $B^\text{eq}(\xx)$. This method is used on Fig~2a-d.


\section{Experiments} \label{app:exp}
\subsection{Experimental system} \label{app:exp_syst}
The experimental system is the one used in Ref.~\cite{Nishiguchi2018}. We used a suspension of
Janus particles of diameter $a = 3.17\pm 0.32~\si{\micro m}$ sandwiched between two ITO electrodes separated by
a spacer of size $H = 50~\si{\micro m}$. 
The particles are suspended in
a sodium chloride solution of concentration $10^{-4}~\si{\mole\per\liter}$.
The use of sodium chloride diminishes the temporal variation of the system, which enables long observation required for calculating pair correlations. 
Furthermore, because the addition of sodium chloride decreases the width of the screening double layer (Debye length) and consequently the strength of the induced hydrodynamic flow~\cite{Squires2006}, both the electrostatic and hydrodynamic interactions between the particles are diminished. This keeps the interactions indiscernible at the frequency range we used here~\cite{Nishiguchi2018}.
For the correlations presented in the main text, we applied
an electric field of frequency $f = 5~$kHz and amplitude $2\cdot 10^6~\si{\volt\pp\per\meter}$ (voltage $10~\si{\volt\pp}$) in the vertical direction. In this range of frequency, the Janus particles move in the direction of the uncoated hemisphere due to the induced-charge electrophoresis~\cite{Nishiguchi2015} and exhibit neither of electrostatic attractive interactions nor strong hydrodynamic interactions~\cite{Nishiguchi2018}. Moreover no polar order is observed even at high density~\cite{Nishiguchi2015}; these experimental optimization justifies the theoretical modeling by Active Brownian Particles as a first approximation.

Videos of the system were captured by a CMOS camera (Baumer, LXG-80, $3000\times 2400$ pixels, 8 bit grayscale) at the framerate 10 fps mounted on an inverted microscope (Olympus, IX70) equipped with a 40$\times$ objective lens (LUCPLFLN, NA=0.60).  A green filter is inserted between the sample and the halogen lamp to increase the contrast of the hemispheres of the Janus particles.
The acquisition length is 14 minutes (8400 frames).

We append an experimental movie at $10~\si{\volt\pp}$, captured at 10 fps. The field of view of this movie is cropped to $585~\si{px}\times 409~\si{px}$ ($70~\si{\micro m}\times 49~\si{\micro m}$) for visibility. The movie is played at the real speed.

\subsection{Image analysis} \label{app:exp_img}
Particles were detected using the Hough Circle Transform algorithm implemented in the OpenCV
library~\cite{opencv2019}. The positions of the particles are the centers of the circles.
For each detected circle, we compute the center of mass of the pixels within it (the weights are the pixels' values)~\cite{Iwasawa2020}. The orientation of the particle is
defined as the direction of the vector between the center of mass and the center of the circle.
We will later give an estimate of the precision of this measure.

\subsection{Experimental parameters} \label{app:exp_params}
We now estimate the parameters of our experimental systems.

\paragraph{Particle diameter.} 
We captured the images with the resolution $0.12~\si{\micro m. px^{-1}}$. In the obtained images of  the particle with the diameter
$a=3.17~\si{\micro m}$, the particle diameter appears $a\simeq 26~\si{px}$.

\paragraph{Density.}  There are on average 487 particles in a $3000\times 2400$ pixels ($\sim 360~\si{\micro m}\times 290~\si{\micro m}$) image. This gives the number density $\rho \simeq \frac{487}{360~\si{\micro m}\times 290~\si{\micro m}}\simeq 0.0047~\si{\micro m^{-2}}$ and the area fraction $\phi = \rho \pi (a/2)^2 \simeq 0.04$.

\paragraph{Velocity.} We use Trackpy~\cite{Trackpy} to obtain the trajectories.
The instantaneous velocities can be obtained by applying a Savitzky-Golay filter~\cite{scipy}.
From the instantaneous velocities, we find an average velocity $U = 56\pm 7~\si{px.s^{-1}}$ (standard deviation given for different particles, see Fig.~\ref{smfig:exp}~(a)). Thus, $U \simeq 6.7~\si{\micro m.s^{-1}}$, that is to say of the order of two particle diameters per second.

Alternatively, one can compute the mean square displacement as a function of time (Fig.~\ref{smfig:exp} (b)). At the lowest order in time, $\langle\Delta x^2\rangle = (U\Delta t)^2$. We find $U \simeq 55~\si{px.s^{-1}} \simeq 6.6~\si{\micro m.s^{-1}}$. This is consistent with the previous result.

\paragraph{Rotational diffusion.}
To measure the rotational diffusion coefficient $D_r$, we compute the mean square angular displacement (MSAD) as a function of time.
At short time, we expect
\begin{equation}
 \langle\Delta\theta^2\rangle = 2\theta_\text{err}^2 + 2 D_r\Delta t, \label{eq:RotationalDiffusion}
\end{equation}
where $\theta_\text{err}$ is the error made in the detection of the orientation of a given particle on a given frame.
At times larger than the mean free time $\tau_\mathrm{free}\simeq (\frac{1}{\sqrt{\rho}}-a)/U\simeq1.7\;\si{s}$, hydrodynamic or electrostatic interactions may affect the orientation of the particles during collisions, resulting in a larger angular displacement.
The MSAD is plotted on Fig.~\ref{smfig:exp}(c); fitting the MSAD with Eq.~(\ref{eq:RotationalDiffusion}) for $0\leq\Delta t\leq 1.5~\si{\second}$ leads to $D_r=0.123\pm 0.004~\si{\second^{-1}}$ and $\theta_\text{err}=8.7\pm 0.3\si{\degree}$.

\begin{figure}
	\begin{center}
		\includegraphics[scale=1]{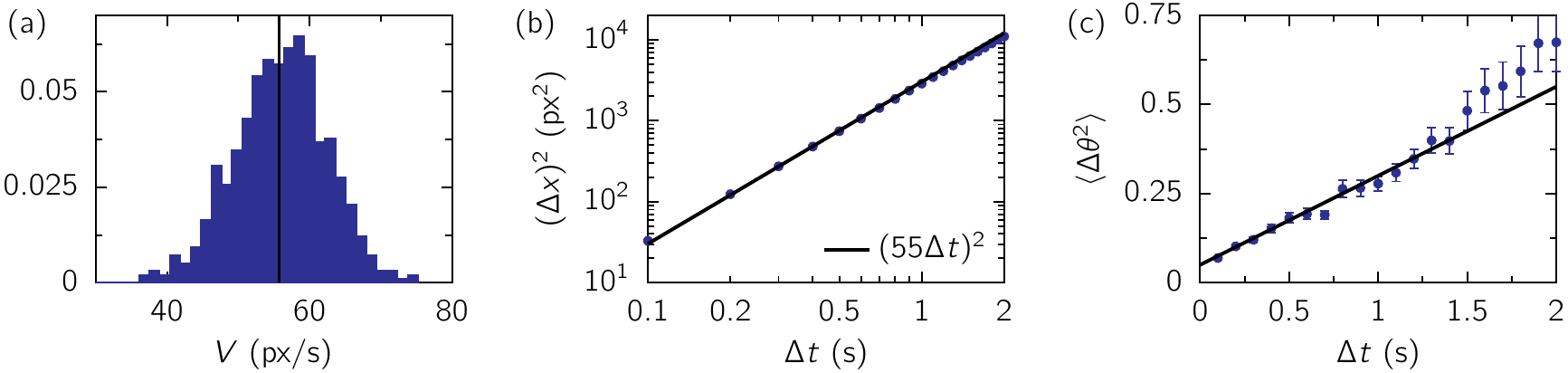}
	\end{center}
	\caption{Determination of the experimental parameters. 670 particles are tracked for 35 frames on average (minimum: 20 frames). (a) Histogram of the average velocities of the particles, the vertical black line is the mean. (b) Mean square displacement. The estimated error bars are smaller than the symbol size. 
	(c) Mean square angular displacement. Error bars represent standard errors estimated from the distribution of $\Delta\theta^2$ for each given $\Delta t$, the solid line is a linear fit over $0\leq\Delta t\leq 1.5~\si{\second}$.
}
	\label{smfig:exp}
\end{figure}

\paragraph{Translational diffusion.} This is the hardest quantity to evaluate.
We give an estimate based on the Stokes-Einstein relation,
\begin{equation}
 D = \frac{k_B T}{6\pi\eta (a/2) \beta},
\end{equation}
$\eta = 1.0\cdot 10^{-3}~\si{Pa.s^{-1}}$ is the viscosity of water, $T\simeq 300~\si{K}$ is the temperature, and $\beta$ is a correction factor due to the proximity of the bottom electrode.
Assuming that Faxen's law (Eq. (7-4.28) of Ref~\cite{Happel1983}) is valid for $h$ (distance to the wall) of the order of $a$, we obtain $\beta\simeq 3$ (one checks that the thermal fluctuations are negligible).
At the end of the day,
\begin{equation}
D \simeq \frac{300\times 1.38\cdot 10^{-23}}{6\pi\times 1.0\cdot 10^{-3}\times 1.585\cdot 10^{-6} \times 3} \simeq 0.05~\si{\micro m^2.s^{-1}}
 \simeq (0.07a)^2/\text{s}.
\end{equation}





\subsection{Correlations}  \label{app:exp_correl}

Once we have detected the positions and the orientations, we consider a given frame.
For each particle far from the edges of the image, we consider every other particle and compute its position in the reference frame of the orientation of the first particle.
We put the result in bins of size $\Delta x = \Delta y = 0.1a$. After processing all the frames and normalizing the bins, we obtain the correlation (Fig~4(b) of the main text).

\subsection{Variation of electric field} \label{app:exp_Efield}
We varied the strength of the electric field in the experiments setting the voltage to $6~\si{\volt\pp}$ and $8~\si{\volt\pp}$ in addition to the $10~\si{\volt\pp}$ data that we report in the article.
The velocity $U$ of the particles is known to scale as the square of the strength of the electric field~\cite{Nishiguchi2015},
thus the P\'eclet number ranges from $\Pe=32$ ($6~\si{\volt\pp}$) to $\Pe=90$ ($10~\si{\volt\pp}$).
The experimental correlations $B(\rr)$ are show on Fig.~\ref{smfig:correl_exp_all}. In all three cases the qualitative behavior is the same: the depletion is maximal in two wings behind the particle.
Experiments with $10~\si{\volt\pp}$ have been performed first, and when the system ages particles start to stick to the substrate, explaining the appearance of positive concentric circles in the correlations for $6$ and $8~\si{\volt\pp}$.

\begin{figure}
\begin{center}
\includegraphics[scale=1]{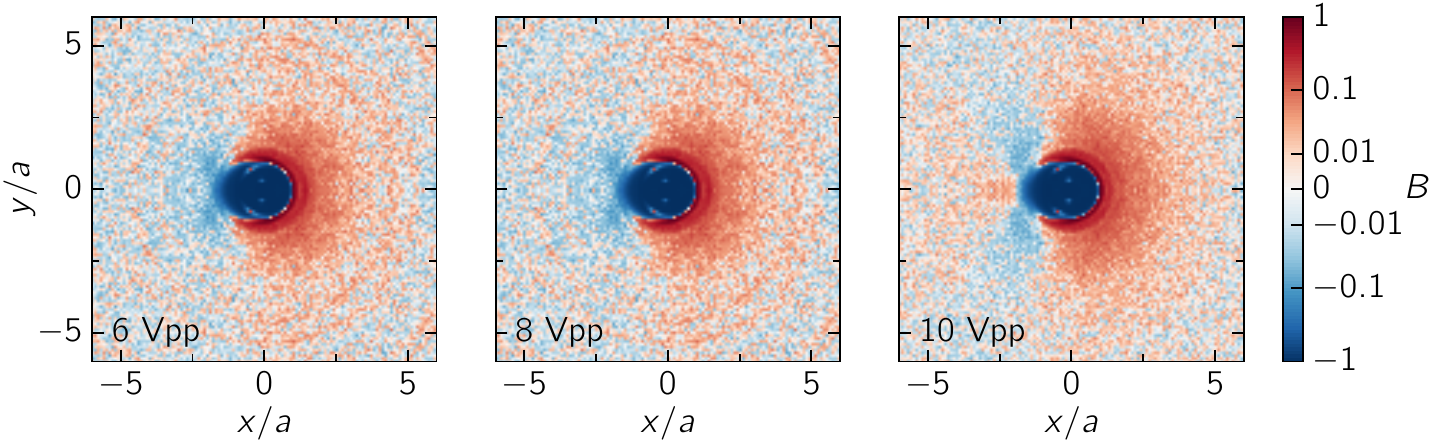}
\end{center}
\caption{Experimental pair correlations for different values of the voltage: $6~\si{\volt\pp}$, $8~\si{\volt\pp}$ and $10~\si{\volt\pp}$ (left to right). The amplitudes of the electric field are $1.2\cdot 10^8$, $1.6\cdot 10^8$ and $2\cdot 10^8~\si{\volt\pp\per\meter}$.}
\label{smfig:correl_exp_all}
\end{figure}
\end{widetext}

%

\end{document}